\documentclass[11pt,a4paper]{article}

\usepackage{a4wide}
\usepackage{latexsym}
\usepackage{epsf}
\usepackage{amssymb}
\usepackage{pst-node}
\usepackage[]{graphicx}

\makeatletter
\@addtoreset{equation}{section}
\makeatother

\begin{document}

\thispagestyle{empty}
\begin{flushright}
\small
IFT-UAM/CSIC-06-01\\
{\bf hep-th/0601003}\\
January $1$st, $2006$
\normalsize
\end{flushright}

\vspace{2cm}

\begin{center}

%title
  {\Large {\bf Supersymmetry and the Supergravity
      Landscape}}\footnote{Enlarged version of the talks given at the
    Pomeranian Workshop on Cosmology and Fundamental Physics (Pobierowo,
    Poland) and the 2005 Spanish Relativity Meeting (in Oviedo, Spain)}

\vspace{2.5cm}

{\large Tom\'as Ort\'{\i}n}

\vspace{1cm}

{\it Instituto de F\'{\i}sica Te\'orica UAM/CSIC\\
  Facultad de Ciencias C-XVI,
  C.U. Cantoblanco,  E-28049-Madrid, Spain}\\
E-mail: {\tt Tomas.Ortin@cern.ch}

\vspace{3cm}

%%%%%%%%%%%%%%%%%%%%%%%%%%%%%%%%%%%%%%%%%%%%%%%%%%%%%%%%%%%%%%%%%%%%%%

{\bf Abstract}

\end{center}

\begin{quotation}

\small
  In the recent times a lot of effort has been devoted to improve our
  knowledge about the space of string theory vacua (``the landscape'')
  to find statistical grounds to justify how and why the theory
  selects its vacuum. Particularly interesting are those vacua that
  preserve some supersymmetry, which are always supersymmetric
  solutions of some supergravity theory. After an general introduction
  to how the pursuit of unification has lead to the vacuum selection
  problem, we are going to review some recent results on the problem
  of finding all the supersymmetric solutions of a supergravity
  theory applied to the $N=4,d=4$ supergravity case.
\end{quotation}

\newpage
%%%%%%%%%%%%%%%%%%%%%%%%%%%%%%%%%%%%%%%%%%%%%%%%%%%%%%%%%%%%%%%%%%%%%%
%%%%%%%%%%%%%%%%%%%%%%%%%%%%%%%%%%%%%%%%%%%%%%%%%%%%%%%%%%%%%%%%%%%%%%
%%%%%%%%%%%%%%%%%%%%%%%%%%%%%%%%%%%%%%%%%%%%%%%%%%%%%%%%%%%%%%%%%%%%%%
%%%%%%%%%%%%%%%%%%%%%%%%%%%%%%%%%%%%%%%%%%%%%%%%%%%%%%%%%%%%%%%%%%%%%%
%%%%%%%%%%%%%%%%%%%%%%%%%%%%%%%%%%%%%%%%%%%%%%%%%%%%%%%%%%%%%%%%%%%%%%

\section{Introduction: Unification and the Landscape}

Unification has been one of the most fruitful guiding principles in
our search for the fundamental components and forces of the Universe.
It is, however, more than just a wish or a prejudice that has produced
important results for a while: it is indeed a logical necessity for
the human mind to understand the Universe: the history of Physics
could be written as the history of the process of unification of many
different concepts, entities and phenomena into an ever smaller and
more fundamental number of them. However, it was only in later times
that we realized what we were doing and started doing it consciously,
setting explicitly the unification of all forces and particles as our
major goal.

It is this (sometimes feverish) pursuit of unification that has lead
us to the vacuum selection problem in Superstring Theory and similar
unification schemes that include gravity. If unification is a major
goal, then, the vacuum selection problem is a major problem of
Superstring Theory, perhaps the most important one.

In order to get some perspective over this problem we are going to review
several instances of unification in Physics. We could go back to Archimedes or
Newton but we will content ourselves with the \textit{classical} period of
unification that starts with Faraday and Maxwell, showing also that the
process of unification underlies all the main advances in Theoretical Physics
and is, in particular, strongly related to the symmetry principles on which
many of our theories are based.

\begin{enumerate}

%%%%%%%%%%%%%%%%%%%%%%%%%%%%%%%%%%%%%%%%%%%%%%%%%%%%%%%%%%%%%%%%%%%%%%
  
\item ${\rm \bf Electricity}\bigoplus{\rm \bf Magnetism}
\stackrel{{\rm \bf Faraday,
      Maxwell}}{\Longrightarrow} {\rm \bf Electromagnetism}$
  
\begin{displaymath}
 \vec{E},  \vec{B} \,\, \longrightarrow \,\, (F_{\mu\nu})\equiv
\left(
  \begin{array}{c|c}
0 & -{\vec{E}}^{T} \\
\hline
{\vec{E}} & {}^{\star}{\vec{B}} \\
  \end{array}
\right)\, .
\end{displaymath}

The unification of electricity and magnetism into a single interaction is the
first paradigm of modern unification of interactions: the unification requires
(or produces) a bigger group of symmetry because the equations of each field
were invariant only under the Galilean group and the full set of Maxwell's
equations are invariant under the Poincar\'e group. This had to be so: if two
interactions are different manifestations of a single interaction, there must
exist transformations that do not change the equations of the theory and
transform one interaction into the other.
 
Had the Special Theory of Relativity been proposed before Maxwell's equations,
the latter could have been discovered by imposing Poincar\'e invariance on the
incomplete equations of electricity and magnetism. However, the importance of
symmetry principles was discovered much later.

Observe that the Principle of (Special) Relativity applied to Newtonian
gravity implies the existence of gravitomagnetism and the combination of both
into a single relativistic field of interaction. This interaction is not yet
General Relativity, but contains its seeds.

\vspace{.2cm}
%%%%%%%%%%%%%%%%%%%%%%%%%%%%%%%%%%%%%%%%%%%%%%%%%%%%%%%%%%%%%%%%%%%%%%

\item ${\rm \bf Space}\bigoplus{\rm \bf Time}
\stackrel{{\rm \bf Einstein, Minkowski}}{\Longrightarrow}  {\rm \bf Spacetime}$

\begin{displaymath}
t,\,\,  \vec{x}
\,\,\longrightarrow\,\, (x^{\mu})\equiv (c t, \vec{x})\, .    
\end{displaymath}

This is an example of unification of fundamental concepts (not interactions),
although it is strongly related to our previous example because the increase
in symmetry (from Galileo to Poincar\'e) is the same and the underlying
mechanism is similar (if space and time are different aspects of spacetime,
there must be transformations that take space into time and vice-versa). 
It is important to observe that the new symmetry is only apparent at high
speeds, but it is never broken.

\vspace{.2cm}
%%%%%%%%%%%%%%%%%%%%%%%%%%%%%%%%%%%%%%%%%%%%%%%%%%%%%%%%%%%%%%%%%%%%%%

\item  ${\rm\bf  Waves}\bigoplus{\rm \bf Particles}
\stackrel{{\rm \bf de Broglie}}{\Longrightarrow}
{\rm\bf Quantum\,\,\, particles}$
  
This unification of two entities always believed to be distinct is required
(and led to) Quantum Mechanics. It is, to this day, mysterious, perhaps
because it is different from the other instances of unification: in this case
there seems to be no underlying symmetry group transforming particles into
waves and vice-versa.

\vspace{.2cm}
%%%%%%%%%%%%%%%%%%%%%%%%%%%%%%%%%%%%%%%%%%%%%%%%%%%%%%%%%%%%%%%%%%%%%%

\item  ${\rm\bf  Gravity\,\,\,     (GR)}\bigoplus{\rm\bf 
    Electromagnetism}\stackrel{{\rm\bf  Kaluza,\,\, Klein,\,\,
      Einstein}}{\Longrightarrow}
  {\rm \bf Higher-dimensional\,\,\, gravity}$

\begin{displaymath}
g_{\mu\nu},  A_{\mu}
\,\, \longrightarrow \,\,
(\hat{g}_{\hat{\mu}\hat{\nu}})\equiv
\left(
  \begin{array}{c|c}
k^{2} & A_{\nu} \\
\hline
A_{\mu} & g_{\mu\nu} \\
  \end{array}
\right)
\end{displaymath}
  
This attempt was unsuccessful (it was, may be, too early) but introduced many
new ideas that have stayed around until now.  In this theory there is also an
increase of symmetry, but the scheme is more complicated: the vacuum of the
theory (in modern parlance) could be 5-dimensional Minkowski spacetime,
invariant under the 5-dimensional Poincar\'e group but this symmetry is
spontaneously broken (again in modern parlance) to the 4-dimensional
Poincar\'e group times $U(1)$ due to the (completely arbitrary) choice of
vacuum (4-dimensional Minkowski spacetime times a circle). General covariance
implies that these symmetries are local in the resulting effective theory, a
fact that can be formulated as the \textit{Kaluza-Klein Principle}:

\begin{quote}
\textsl{Global invariances of the vacuum are local invariances of the theory}.
\end{quote}

An, originally unwanted, feature of the theory is that a new massless field is
predicted: the \textit{Kaluza-Klein scalar} (or \textit{radion}) $k$. Its
v.e.v., related to the radius of the internal circle, can also be fixed
arbitrarily because there is no potential for this scalar. Fixing
(\textit{stabilizing}) the v.e.v.~of scalars such as $k$ that determine the
size and shape of part of the vacuum spacetime (generically known as
\textit{moduli}) is nowadays known as the \textit{moduli problem}. Explaining
why the vacuum should be 4-dimensional Minkowski spacetime times a circle of
all the possible classical solutions of 5-dimensional General Relativity is
the simplest version of the \textit{vacuum selection problem}.

\vspace{.2cm}
%%%%%%%%%%%%%%%%%%%%%%%%%%%%%%%%%%%%%%%%%%%%%%%%%%%%%%%%%%%%%%%%%%%%%%

\item  ${\rm \bf Quantum\,\,\, Mechanics}\bigoplus
{\rm \bf Relativistic\,\,\, Field\,\,\, Theory}
  \stackrel{{\rm\bf  Many\,\,\, people}}{\Longrightarrow}
  {\rm\bf  QFT}$

  A difficult but fruitful marriage.

\vspace{.2cm}
%%%%%%%%%%%%%%%%%%%%%%%%%%%%%%%%%%%%%%%%%%%%%%%%%%%%%%%%%%%%%%%%%%%%%%

\item ${\rm \bf Weak\,\,\, interactions}\bigoplus{\rm \bf  Electromagnetism}
  \stackrel{{\rm \bf Glashow,\,\,\, Salam,\,\,\, Weinberg}}{\Longrightarrow}
  {\rm \bf EW\,\,\, interaction}$

In this case, two Relativistic QFTs are unified. 

\begin{itemize}
\item Unification is achieved by an increase of \textit{local}
  (Yang-Mills-type) symmetry, from $U(1)$ to $SU(2)\times U(1)$.
\item The symmetry is \textit{spontaneously broken} by the Higgs mechanism:
  choice of vacuum by energetic reasons (minimization of the {\it ad hoc}
  Higgs potential). (This is the main difference with Kaluza-Klein and other
  theories including gravity in which different vacua are associated to
  different spacetimes and, therefore, different definitions of energy that
  cannot be compared.)

\item The spontaneous breaking of the symmetry renders the
  model renormalizable.

\item The symmetry is restored at high energies.

\item New massive particles are predicted associated to the enhanced symmetry
  (gauge bosons, found) and a new massless spin-0 particle is also predicted
  (Higgs boson, not yet found).

\end{itemize}

This model, part of the Standard Model of Particle Physics, has had an
extraordinary success and most unification schemes of relativistic QFTs have
followed the same pattern. In particular

\vspace{.2cm}
%%%%%%%%%%%%%%%%%%%%%%%%%%%%%%%%%%%%%%%%%%%%%%%%%%%%%%%%%%%%%%%%%%%%%%

\item  ${\rm \bf Electroweak\,\,\,  interaction}
\bigoplus{\rm \bf Strong\,\,\, interactions}
  \stackrel{{\rm \bf Many\,\,\, people...}}{\Longrightarrow}
  {\rm \bf Grand\,\,\, Unified\,\,\, Theory}$
  
  This is an unsuccessful generalization of the electroweak unification scheme
  based on a semisimple gauge group ($SO(10), SU(5),\cdots$) spontaneously
  broken by a generalized Higgs mechanism to $SU(3)\times U(1)$. There are two
  main problems:
  \begin{itemize}
  \item New massive and massless particles predicted may mediate proton
      disintegration (not observed).
  \item Unification of coupling constants should occur at the energy at
    which the symmetry is restored, but this does not seems to work.
\end{itemize}

\vspace{.2cm}
%%%%%%%%%%%%%%%%%%%%%%%%%%%%%%%%%%%%%%%%%%%%%%%%%%%%%%%%%%%%%%%%%%%%%%

\item  ${\rm \bf Bosons}\bigoplus{\rm \bf Fermions}
  \stackrel{{\rm \bf Golfand, Likhtman, Volkov, Akulov, Soroka,
      Wess\, and\, Zumino}}{\Longrightarrow}   {\rm \bf Superfields}$
  
  This is a new kind of unification based in an increase of (global spacetime)
  symmetry to \textit{supersymmetry}, which should also be spontaneously
  broken by a yet unknown super-Higgs mechanism. It has many interesting
  properties:

  \begin{itemize}
        
  \item It is the most general extension of the Poincar\'e and {Yang-Mills}
    symmetries of the S-matrix (Haag-Lopuszanski-Sohnius theorem).

  \item This new symmetry can be combined with Yang-Mills-type symmetries
    (super-Yang-Mills theories) and with GUT models in which, in some cases,
    unification of coupling constants can be achieved.
    
  \item It can also be combined with g.c.t.'s, making it local
    (\textit{supergravity} theories). We can have supergravity theories with
    Yang-Mills fields etc., but in most of these theories gravity is not
    unified with the other interactions since they belong to different
    supermultiplets.
    
  \item However, \textit{extended} ($N>1$) supergravities contain in the same
    supermultiplet of the graviton additional bosonic fields that may describe
    the other interactions. In this scheme all interactions would be described
    in a truly unified way.
    
    These extended supergravities can in general be obtained from
    compactification of simpler higher-dimensional supergravities. It was also
    discovered that many $N=1$ supergravities coupled to Yang-Mills fields
    could also be obtained in the same way, by a careful choice of compact
    manifold (i.e.~Kaluza-Klein vacuum). This lead to a new brand of unified
    theories which could describe everything (\textit{Theories of
      Everything}). The first of these is
\end{itemize}

\vspace{.2cm}
%%%%%%%%%%%%%%%%%%%%%%%%%%%%%%%%%%%%%%%%%%%%%%%%%%%%%%%%%%%%%%%%%%%%%%

\item \textbf{Kaluza-Klein Supergravity  \cite{Duff:1986hr,Appelquist:1987nr}} 
  
  It is a combination of the Kaluza-Klein theories with supersymmetry. Now, a
  Kaluza-Klein vacuum is (arbitrarily) chosen that breaks spontaneously part
  of the (super)symmetries of the ``original'' vacuum (Minkowski spacetime for
  Poincar\'e supergravities and anti-De Sitter spacetime for aDS
  supergravities). Now, the rule of the game, the \textit{supersymmetric
    Kaluza-Klein Principle}, is

\begin{quote}
\textsl{Global (super)symmetries of the vacuum are local (super)symmetries of
  the compactified theory}.
\end{quote}

In general, the theories were based on compactifications of $N=1,d=11$
supergravity \cite{Cremmer:1978km}, the unique supergravity that can be
constructed in the highest dimension in which a consistent supergravity can be
constructed. It can accommodate the bosonic part of the Standard Model with
minimal supersymmetry. However, these theories are anomalous and it is
impossible to obtain the chiral structure of the Standard Model by
compactification on \textit{smooth} manifolds \cite{Witten:1983ux}.  The
vacuum of these theories was arbitrarily chosen to recover the Standard Model.
The arbitrariness in the choice of vacuum replaces that of the choice of Higgs
field and potential (and gauge interactions, dimensionality...). This makes
these theories, conceptually, far superior, but raises to a very prominent
place the vacuum selection problem.

These problems and the advent of String Theory, in particular the Heterotic
Superstring \cite{Gross:1985fr}, which is anomaly-free and has chiral
fermions, killed these theories, although they have been resurrected again by
the same theory that killed them.

\vspace{.2cm}
%%%%%%%%%%%%%%%%%%%%%%%%%%%%%%%%%%%%%%%%%%%%%%%%%%%%%%%%%%%%%%%%%%%%%%
\item   {\bf Superstring Theories}
  
  In these theories, all quantum particles are different vibration states of a
  single physical entity: the superstring. All known interactions could be
  described in this way. At low energies, one recovers an anomaly-free
  supergravity theory.  However, there are still some problems:
  \begin{itemize}
  \item They are 10-dimensional, and require compactification.  At low
    energies we are faced with 10-dimensional Kaluza-Klein supergravity and
    the vacuum selection problem.
    
  \item There are at least five superstring theories: Types~IIA, Type~IIB,
    Heterotic~$SO(32)$, Heterotic~$E(8)\times E(8)$ and Type~I $SO(32)$. Which
    one should be considered?
    
  \item The theory seems to contain other extended objects besides strings:
    \textit{D-branes} \cite{Polchinski:1995mt}, NSNS-branes...  Why should
    strings be fundamental \cite{Townsend:1995gp}?
\end{itemize}

The answer to the last two questions lies on the \textit{dualities} that
related the different superstring theories and the different extended objects
that occur in them \cite{Schwarz:1996bh}.  Dualities are transformations that
relate different theories: their spectra, interactions and coupling constants.
Their existence allows the mapping of all scattering amplitudes of one theory
into those of the other theory and vice-versa. In some cases, the mapping
relates the coupling constant of one theory with the inverse coupling constant
of the other theory and we talk about the non-perturbative S~dualities. In
other cases the non-trivial mapping affects only geometrical data of the
compactification (moduli) and we talk about perturbative T~dualities. These
are characteristic of String Theories.

Dualities are, certainly, not symmetries of a single theory. Instead, they can
be seen as symmetries in the space of theories. If two dual theories arise
from two different compactifications (i.e.~choices of vacua) of a given String
Theory, then dualities can be seen as symmetries in the space of vacua. The
extrapolation of this fact to the cases in which the theories are not known to
originate from the same theory by different choices of vacuum is the basis of
\textit{M~Theory}. 

\vspace{.2cm}
%%%%%%%%%%%%%%%%%%%%%%%%%%%%%%%%%%%%%%%%%%%%%%%%%%%%%%%%%%%%%%%%%%%%%%

\item $\bigoplus{\rm\bf 
    Superstring~Theories} \stackrel{{\rm \bf 
      Witten~et~al.}}{\Longrightarrow}
  {\rm\bf  M~theory}$
  
  In this (super-) unification scheme, all the superstring theories are
  understood as different duality-related vacua of an unknown theory called
  M~Theory, whose low energy limit is $N=1,d=11$ supergravity, which was
  discovered by Witten \cite{Witten:1995ex} to be related to the
  strong-coupling limit (S~duality) of the low-energy limit of Type~IIA
  Superstring ($N=2A,d=10$ Supergravity).
 
  Now we are back, in a sense, into the old Kaluza-Klein supergravity
  scenario, but all the Supergravity fields have got a String Theory meaning.
  It is amusing to see how, in this scheme, the low-energy limit of the
  Heterotic Superstring , which has chiral fermions, is related to the
  low-energy limit of M~theory: 11-dimensional supergravity, which was
  apparently forbidden by Witten's no-go theorem \cite{Witten:1983ux}. The
  solution to the inconsistency is the use of non-smooth manifolds (orbifolds)
  \cite{Horava:1995qa}, evading one the hypothesis of the theorem. This could
  have been done many years earlier, but, without Superstring Theory
  underlying the Supergravity theory other problems such as anomalies may
  never have been solved.

\end{enumerate}

%%%%%%%%%%%%%%%%%%%%%%%%%%%%%%%%%%%%%%%%%%%%%%%%%%%%%%%%%%%%%%%%%%%%%%

The unification scheme proposed by M~theory is very attractive and could
satisfy all our desires for unification: all particles and interactions may be
explained in a unified way. Further, we no longer have different Superstring
Theories to choose from. All the arbitrariness we had have disappeared, but
only to be replaced by the arbitrariness in the choice of vacuum. Now there is
only one theory and everything depends on that. But the theory seems to have
nothing to tell us yet about how it chooses the vacuum and why our Universe is
as we see it.

It has to be mentioned that, nowadays, we ask much more from a good candidate
to \textit{the} vacuum of our theory: it is not enough (but it is, certainly,
a good starting point) that it gives the Standard Model of Particle Physics,
but it should also explain the evolution of our Universe, that is, according
to the most extended prejudices, it should give rise to an inflationary era
and explain, in a fundamental way, dark energy.

With respect to this problem, there have been two main directions of work:

\begin{itemize}
\item Finding phenomenologically viable vacua (in the Particle Physics,
  e.g.~\cite{Ibanez:2001nd} and/or cosmological, e.g.~\cite{Kachru:2003sx}
  sense).
\item Find a vacuum-selection mechanism.
\end{itemize}

There has been no real progress in the second direction for many years.

The failure to solve the vacuum selection problem through some
dynamical mechanism has favored recently a purely statistical approach
in which one first has to explore and chart (``classify'') the space
of vacua a.k.a.~\textit{Landscape}.  In this approach, our Universe is
the way it is because the probability of this kind of Universe is
overwhelming.  Of course, this way of thinking can be combined with
different forms of the Anthropic Principle.

Charting the superstring landscape is a very difficult problem and
some simplifications have been suggested: for instance, one could
consider all supersymmetric String Theory vacua, which correspond to
different kinds of supergravities \cite{Susskind:2003kw} or only the
vacua with 4-dimensional Poincar\'e symmetry and a Calabi-Yau internal
space, which correspond to $N=1,d=4$ supergravities and give
Standard-Model-like theories \cite{Douglas:2003um}. One could also
consider, as proposed by Van Proeyen \cite{kn:VP}, all possible
supergravities, even if the stringy origin of many of them is unknown
(the \textit{supergravity landscape}).

In this talk, which is based on
Refs.~\cite{Kallosh:1993wx,Bellorin:2005hy,Bellorin:2005zc}, we are going to
review some recent general results on the classification of supersymmetric
String Theory vacua and new techniques that can be used to find them,
presenting some particular results on the classification of the supersymmetric
vacua of the toroidally compactified Heterotic String Theory ($N=4,d=4$
SUGRA). First, we are going to define what is a supersymmetric configuration
and its symmetry superalgebra, describing some useful special identities that
they satisfy (\textit{Killing spinor identities}).  Then we will move on to
define the problem of finding all the supersymmetric configurations of a given
supergravity theory \textit{Tod's problem} and we will explain the strategy to
solve it in most (4-dimensional) cases. Finally, we will consider the case of
$N=4,d=4$ supergravity.

%%%%%%%%%%%%%%%%%%%%%%%%%%%%%%%%%%%%%%%%%%%%%%%%%%%%%%%%%%%%%%%%%%%%%%
%%%%%%%%%%%%%%%%%%%%%%%%%%%%%%%%%%%%%%%%%%%%%%%%%%%%%%%%%%%%%%%%%%%%%%
%%%%%%%%%%%%%%%%%%%%%%%%%%%%%%%%%%%%%%%%%%%%%%%%%%%%%%%%%%%%%%%%%%%%%%
%%%%%%%%%%%%%%%%%%%%%%%%%%%%%%%%%%%%%%%%%%%%%%%%%%%%%%%%%%%%%%%%%%%%%%
%%%%%%%%%%%%%%%%%%%%%%%%%%%%%%%%%%%%%%%%%%%%%%%%%%%%%%%%%%%%%%%%%%%%%%
%%%%%%%%%%%%%%%%%%%%%%%%%%%%%%%%%%%%%%%%%%%%%%%%%%%%%%%%%%%%%%%%%%%%%%
%%%%%%%%%%%%%%%%%%%%%%%%%%%%%%%%%%%%%%%%%%%%%%%%%%%%%%%%%%%%%%%%%%%%%%
%%%%%%%%%%%%%%%%%%%%%%%%%%%%%%%%%%%%%%%%%%%%%%%%%%%%%%%%%%%%%%%%%%%%%%
%%%%%%%%%%%%%%%%%%%%%%%%%%%%%%%%%%%%%%%%%%%%%%%%%%%%%%%%%%%%%%%%%%%%%%

\section{Supersymmetric configurations and solutions}

Supersymmetric configurations\footnote{It will be very important for our
  discussion to distinguish between general field configurations and
  (classical) solutions of a given theory. General field configurations may or
  may not satisfy the classical equations of motion and, therefore, may or may
  not be classical solutions. As we are going to see, supersymmetry does not
  ensure that the equations of motion are satisfied.} (a.k.a.~configurations
with residual or unbroken or preserved supersymmetry) are classical bosonic
configurations of supergravity (SUGRA) theories which are invariant under some
supersymmetry transformations.  Let us see what this definition implies.

Generically, the supersymmetry transformations take, schematically,
the form

\begin{equation}
\label{eq:genericsusyrules}
  \delta_{\epsilon} \phi^{b}
 \sim \bar{\epsilon} \phi^{f}\, ,
\hspace{1cm}
  \delta_{\epsilon} \phi^{f}
 \sim \partial \epsilon
+\phi^{b}\epsilon\, ,
\end{equation}

\noindent
where $\phi^{b}$ stands for bosonic fields (or products of an even
number of fermionic fields) and $\phi^{f}$ for the fermionic fields
(or products of an odd number of fermionic fields) and $\epsilon$ are
the infinitesimal, local, parameters of the supersymmetry
transformations, which are fermionic.

Then, a bosonic configuration (i.e.~a configuration with vanishing fermionic
fields $\phi^{f}=0$) will be invariant under the infinitesimal supersymmetry
transformation generated by the parameter $\epsilon^{\alpha}(x)$ if it
satisfies the {\it Killing spinor equations} (one equation for each
$\phi^{f}$), which have the generic form

\begin{equation}
\label{eq:genericKillingspinorequation}
  \delta_{\epsilon} \phi^{f} \sim \partial \epsilon +\phi^{b} \epsilon=0\, .
\end{equation}

The concept of unbroken supersymmetry is a generalization of the
concept of isometry, an infinitesimal general coordinate
transformation generated by $\xi^{\mu}(x)$ that leaves the metric
$g_{\mu\nu}$ invariant because it satisfies the {\it Killing (vector)
  equation}

\begin{equation}
\label{eq:Killingvectorequation}
  \delta_{\xi} g_{\mu\nu}= 2\nabla_{(\mu}\xi_{\nu)}=0\, .
\end{equation}

\noindent
As it is well known, in this case, to each bosonic symmetry we
associate a generator

\begin{equation}
\xi_{(I)}^{\mu}(x)\rightarrow P_{I}\, ,
\end{equation}

\noindent
of a symmetry algebra 

\begin{equation}
[P_{I},P_{J}]  =f_{IJ}{}^{K}  P_{K}\, ,\,\,\,
\Leftrightarrow [\xi_{(I)},\xi_{(J)}]  =f_{IJ}{}^{K}  \xi_{(K)}\, ,
\end{equation}

\noindent
where the brackets in the right are Lie brackets of vector fields.

In our case, the unbroken supersymmetries are associated to the odd
generators

\begin{equation}
\epsilon_{(n)}^{\alpha}(x)\rightarrow\mathcal{Q}_{n}\, , 
\end{equation}

\noindent
of a superalgebra

\begin{equation}
[\mathcal{Q}_{n},P_{I}]  =f_{nI}{}^{m}\mathcal{Q}_{m}  \, ,
\hspace{1cm}
\{\mathcal{Q}_{n},\mathcal{Q}_{m}\}  =f_{nm}{}^{I}P_{I}\, .  
\end{equation}

The calculation of these commutators and anticommutators is explained in
detail in Refs.~\cite{Ortin:2002qb,Alonso-Alberca:2002dw} and the consistency
of the scheme was proven in \cite{Figueroa-O'Farrill:2004mx}.  According to
the Kaluza-Klein principle we enunciated at the beginning, conveniently
generalized to the supersymmetric case, this global supersymmetry algebra
becomes the algebra of the local symmetries of the field theories constructed
on this field configuration.

Of course, we do not want to construct field theories on just any field
configuration but only on vacua of the theory.  In general, for a field
configuration to be considered a vacuum, we require that it is a classical
solution of the equations of motion of the theory.  Apart from this
requirement, it is not clear what \textit{a priori} characteristics a good
vacuum must have except for classical and quantum stability, which are
difficult to test in general, but which are, under certain conditions,
guaranteed by the presence of unbroken supersymmetry. This is one of the
reasons that makes supersymmetric vacua interesting. We also \textit{prefer}
highly symmetric vacua (such as Minkowski or anti-De Sitter space) since, on
them, we can define a large number of conserved quantities, but it is
uncertain why Nature should have the same prejudices.

Sometimes, when a vacuum solution has a clear (possibly warped) product
structure, we can distinguish internal and spacetime (super-) symmetries and,
if we choose this vacuum, our choice implies spontaneous compactification.

%%%%%%%%%%%%%%%%%%%%%%%%%%%%%%%%%%%%%%%%%%%%%%%%%%%%%%%%%%%%%%%%%%%%%%
%%%%%%%%%%%%%%%%%%%%%%%%%%%%%%%%%%%%%%%%%%%%%%%%%%%%%%%%%%%%%%%%%%%%%%
%%%%%%%%%%%%%%%%%%%%%%%%%%%%%%%%%%%%%%%%%%%%%%%%%%%%%%%%%%%%%%%%%%%%%%

\section{Tod's problem}

This is the problem of finding \textit{all} the supersymmetric bosonic field
configurations, i.e.~all the bosonic field configurations $\phi^{b}$ for which
a SUGRA's Killing spinor equations

\begin{equation}
\left.
\delta_{\epsilon}\phi^{f}
\right|_{\phi^{f}=0}\sim 
\partial \epsilon+\phi^{b}
\epsilon =0\, ,      
\end{equation}

\noindent    
have a solution $\epsilon$, which includes all the possible supersymmetric
vacua and compactifications.

Observe that, as we announced, not all supersymmetric bosonic field
configurations satisfy the classical bosonic equations of motion for which we
use the notations $\left.\frac{\delta S}{\delta \phi^{b}}\right|_{\phi^{f}=0}
\equiv \left.  S_{,b}\right|_{\phi^{f}=0} \equiv \mathcal{E}(\phi^{b})$.
Actually, the bosonic equations of motion of supersymmetric bosonic field
configurations satisfy the so-called {\it Killing spinor identities} (KSIs)
\cite{Kallosh:1993wx,Bellorin:2005hy}that relate different equations of motion
of a supersymmetric theory.  These identities can be derived as follows: The
supersymmetry invariance of the action implies, for arbitrary local
supersymmetry parameters $\epsilon$

\begin{equation}
\delta_{\epsilon}S
=
\int d^{d}x\, (
S_{,b}\, 
\delta_{\epsilon}\phi^{b}   
+ S_{,f}\, 
\delta_{\epsilon}\phi^{f}
)
=0\, .
\end{equation}

\noindent
Taking the functional derivative w.r.t.~the fermions and setting them to zero

\begin{equation}
%% \left( \right.
%% \delta_{{\epsilon}}S
%% \left.\left. \right)_{, f_{1}}
%% \right|_{\phi^{f}=0}
%% =
\left.
\int d^{d}x\, 
\left[
S_{,bf_{1}}\, 
\delta_{\epsilon}\phi^{b}   
+
S_{,b}\, 
(\delta_{\epsilon}\phi^{b})_{,f_{1}}
+ 
S_{,ff_{1}}\, 
\delta_{\epsilon}\phi^{f}
+
S_{,f}\, 
(\delta_{\epsilon}\phi^{f})_{,f_{1}}
\right]
\right|_{\phi^{f}=0}
=0\, ,
\end{equation}

The terms $\left.\delta_{\epsilon}\phi^{b} \right|_{\phi^{f}=0}\, ,\, \left.
  S_{,f} \right|_{\phi^{f}=0}\, ,\, \left.  (\delta_{\epsilon}\phi^{f}
  )_{,f_{1}} \right|_{\phi^{f}=0} $ vanish automatically because they are odd
in fermion fields $\phi^{f}$ and so we are left with

\begin{equation}
\left.\left\{S_{,b}\, 
(\delta_{\epsilon}
\phi^{b})_{,f_{1}}
+S_{,ff_{1}}\, 
\delta_{\epsilon}\phi^{f}
\right\}\right|_{\phi^{f}=0}=0\, .
\end{equation}

This is valid for any fields $\phi^{b}$ and any supersymmetry parameter
$\epsilon$.  For a supersymmetric field configuration $\epsilon$ is a Killing
spinor $\left.  \delta_{\epsilon}\phi^{f} \right|_{\phi^{f}=0}$ and we obtain
the KSIs

\begin{equation}
\label{eq:ksis}
\mathcal{E}(\phi^{b})
\left. (\delta_{\epsilon}
\phi^{b})_{,f_{1}}
\right|_{\phi^{f}=0}
=0\, .
\end{equation}

These non-trivial identities are linear relations between the bosonic
equations of motion and can be used to solve Tod's problem, obtain BPS bounds
etc.  Let's see some examples.

%%%%%%%%%%%%%%%%%%%%%%%%%%%%%%%%%%%%%%%%%%%%%%%%%%%%%%%%%%%%%%%%%%%%%%
%%%%%%%%%%%%%%%%%%%%%%%%%%%%%%%%%%%%%%%%%%%%%%%%%%%%%%%%%%%%%%%%%%%%%%
%%%%%%%%%%%%%%%%%%%%%%%%%%%%%%%%%%%%%%%%%%%%%%%%%%%%%%%%%%%%%%%%%%%%%%

\subsection{Example: $N=1,d=4$ Supergravity.}

This is the simplest supergravity theory. Its field content is
$\{e^{a}{}_{\mu},\psi_{\mu}\}$.  The bosonic action (Einstein-Hilbert's) and
the equations of motion (Einstein's) are

\begin{equation}
\left. S \right|_{\psi_{\mu}=0}
=\int d^{4}x \sqrt{|g|}\, 
R\, , \,\,\, \Rightarrow\,\,\,
\mathcal{E}_{a}{}^{\mu}({e}) \sim G_{a}{}^{\mu}\, .  
\end{equation}

\noindent
The supersymmetry transformations of the graviton and gravitino are

\begin{equation}
\delta_{\epsilon} e^{a}{}_{\mu} =
-i\bar{\epsilon}\gamma^{a}\psi_{\mu}\, ,
\hspace{1cm}
\delta_{\epsilon} \psi_{\mu} =
\nabla_{\mu}\epsilon
=\partial_{\mu}\epsilon 
-{\textstyle\frac{1}{4}}\omega_{\mu}{}^{ab}
\gamma_{ab}\epsilon\, .
\end{equation}

\noindent
The KSIs can be readily computed from the general formula Eq.~(\ref{eq:ksis})
and simplified

\begin{equation}
-i\bar{\epsilon} \gamma^{a}  
G_{a}{}^{\mu}=0\, ,\,\,\,\, 
\Rightarrow 
R=0\, ,\,\,\,
-i\bar{\epsilon} \gamma^{a}  
R_{a}{}^{\mu}=0\, .
\end{equation}

On the other hand, in trying to solve the Killing spinor equations (KSEs)
which, here, take the form $\delta_{\epsilon} \psi_{\mu}=
\nabla_{\mu}\epsilon=0$, we can consider first their integrability conditions:

\begin{equation}
[\nabla_{\mu},\nabla_{\nu}] \epsilon=
-{\textstyle\frac{1}{4}} R_{\mu\nu}{}^{ab}\gamma_{ab}
\epsilon =0\, ,\,\,\,
\Rightarrow 
R^{\mu}{}_{a}\gamma^{a} \epsilon=0\, .
\end{equation}

Thus, at least in the case, the KSIs are contained in the integrability
conditions.  We will see later how to obtain more information from these
identities.

%%%%%%%%%%%%%%%%%%%%%%%%%%%%%%%%%%%%%%%%%%%%%%%%%%%%%%%%%%%%%%%%%%%%%%
%%%%%%%%%%%%%%%%%%%%%%%%%%%%%%%%%%%%%%%%%%%%%%%%%%%%%%%%%%%%%%%%%%%%%%
%%%%%%%%%%%%%%%%%%%%%%%%%%%%%%%%%%%%%%%%%%%%%%%%%%%%%%%%%%%%%%%%%%%%%%

\subsection{Example: $N=2,d=4$ Supergravity.}

This is the next simplest supergravity theory, if we do not consider adding
matter supermultiplets to the $N=1$ theory.  Its field content is
$\{e^{a}{}_{\mu},A_{\mu}, \psi_{\mu}\}$ (but now $ \psi_{\mu}$ is a Dirac
spinor, instead of a Majorana spinor as in the $N=1$ case).  The bosonic
action (Einstein-Maxwell's) and the equations of motion (Einstein's and
Maxwell's) are

\begin{equation}
\left. S \right|_{\psi_{\mu}=0}
=\int d^{4}x \sqrt{|g|}\, 
\left[R
-{\textstyle\frac{1}{4}}F^{2}\right]\, , 
\,\,\, \Rightarrow\,\,\,
\left\{ 
  \begin{array}{rcl}
\mathcal{E}_{a}{}^{\mu}(e) & = & 
-2 \{ G_{a}{}^{\mu}
-\frac{1}{2} T_{a}{}^{\mu} 
\}\, , \\ 
& & \\
\mathcal{E}^{\mu}(A) & = &  
\nabla_{\alpha} F^{\alpha\mu}\, . \\
  \end{array}
\right.
\end{equation}

\noindent
The supersymmetry transformations are

\begin{equation}
\delta_{\epsilon} e^{a}{}_{\mu} =
-i\bar{\epsilon}\gamma^{a}\psi_{\mu}
+{\rm c.c.}\, ,
\hspace{.5cm}
\delta_{\epsilon} A_{\mu} =
 -2i\bar{\epsilon}\psi_{\mu} 
+{\rm c.c.}\, .
\hspace{.5cm}
\delta_{\epsilon} \psi_{\mu} =
\nabla_{\mu}\epsilon
-{\textstyle\frac{1}{8}}F^{ab}
\gamma_{ab}\epsilon\equiv 
\tilde{\cal D}_{\mu}\epsilon\, .
\end{equation}

\noindent
Using the bosonic fields supersymmetry transformations, we find that the KSIs
take the form

\begin{equation} 
\bar{\epsilon}\{ \mathcal{E}_{a}{}^{\mu} (e) 
\gamma^{a} +2\mathcal{E}^{\mu}(A)\} = 0\, . 
\end{equation}

\noindent
On the other hand, the integrability conditions of the KSEs $\delta_{\epsilon}
\psi_{\mu} = \tilde{\cal D}_{\mu}\epsilon=0$ are

\begin{equation}
  \begin{array}{l}
[\tilde{\cal D}_{\mu},\tilde{\cal D}_{\nu}]\epsilon
=
-{\textstyle\frac{1}{4}} 
\left\{
\left[
R_{\mu\nu}{}^{ab}
-e^{a}{}_{[\mu}T_{\nu]}{}^{b}
\right]\gamma_{ab}
+\nabla^{a} 
\left( F_{\mu\nu} 
+{}^{\star}F_{\mu\nu} \gamma_{5}\right)\gamma_{a}
\right\} \epsilon=0\, ,\\
\\
\\
\Rightarrow\,\,\,
\{ \mathcal{E}_{a}{}^{\mu} (e) \gamma^{a} 
+2[\mathcal{E}^{\mu}({A}) 
+\mathcal{B}^{\mu}(A) \gamma_{5}] \} 
\epsilon = 0\, .
\end{array}
\end{equation}

In this case we get a more general formula from the integrability conditions,
valid for the case in which the Bianchi identities are not satisfied. When
they are satisfied we recover the KSIs, which is consistent since we have
explicitly used the supersymmetry variations of the vector field in order to
derive them, assuming, then, implicitly, that the Bianchi identities are
satisfied.

The last formula (which we are also going to call KSI) has one important
advantage over the original KSI: it is covariant under the $U(1)$ group of
electric-magnetic duality rotations of the Maxwell and Bianchi identities that
act as chiral rotations of the spinors.

%%%%%%%%%%%%%%%%%%%%%%%%%%%%%%%%%%%%%%%%%%%%%%%%%%%%%%%%%%%%%%%%%%%%%%
%%%%%%%%%%%%%%%%%%%%%%%%%%%%%%%%%%%%%%%%%%%%%%%%%%%%%%%%%%%%%%%%%%%%%%
%%%%%%%%%%%%%%%%%%%%%%%%%%%%%%%%%%%%%%%%%%%%%%%%%%%%%%%%%%%%%%%%%%%%%%

\section{Solving Tod's problem}

In 1983 showed in Ref.~\cite{Tod:1983pm} that in $N=2,d=4$ SUGRA the problem
could be completely solved using just integrability and consistency
conditions.  However, he used the Newman-Penrose formalism, unfamiliar to
most particle physicists and suited only for $d=4$. Thus, there were no
further results until 1995, when Tod, using again the same methods, solved
partially the problem in $N=4,d=4$ SUGRA \cite{Tod:1995jf}.  Then, in 2002,
Gauntlett, Gutowski, Hull, Pakis and Reall proposed to translate the Killing
spinor equation to tensor language and they solved the problem in minimal
$N=1,d=5$ SUGRA \cite{Gauntlett:2002nw}. This opened the gates to new results:
in 2002 the problem was solved in \textit{gauged} minimal $N=1,d=5$ SUGRA
\cite{Gauntlett:2003fk}, in 2003 in minimal $N=(1,0),d=6$ SUGRA
\cite{Gutowski:2003rg,Chamseddine:2003yy} and gauged $N=2,d=4$ SUGRA
\cite{Caldarelli:2003pb}, and in 2004 and 2005 in gauged minimal $N=1,d=5$
SUGRA coupled to Abelian vector multiplets
\cite{Gutowski:2004yv,Gutowski:2005id} and in $N=4,d=4$ SUGRA
\cite{Bellorin:2005zc}, completing the work started by Tod on this theory.

There is by now a well-defined recipe to attack this problem (at least in low
dimensions) starting with only one assumption: the existence of one Killing
spinor $\epsilon$. The recipe consists in the following steps:

\begin{itemize}
  
\item[I] Translate the {Killing spinor} equations and KSIs into tensorial
  equations.
  
  With the Killing spinor $\epsilon$ one can construct scalar, vector, and
  $p$- form bilinears $M\sim \bar{\epsilon} \epsilon\, ,\hspace{.5cm}
  V_{\mu}\sim \bar{\epsilon} \gamma_{\mu} \epsilon\, , \cdots$ that are
  related by Fierz identities. These bilinears satisfy certain equations
  because they are made out of Killing spinors, for instance, if the KSE is of
  the general form

\begin{equation}
\label{eq:Killingeq}
\delta_{\epsilon} \psi_{\mu} =
\tilde{\cal D}_{\mu}\epsilon=
[\nabla_{\mu}+\Omega_{\mu}]
\epsilon=0\, ,\,\,\,
\Rightarrow\,\,
\nabla_{\mu} M +2 \Omega_{\mu}M=0\, ,  
\end{equation}

The set of all such equations for the bilinears should be equivalent to the
original spinorial equation or at least it should contain most of the
information contained in it (but, certainly, not all of it).

\item[II] One of the vector bilinears (say $V_{\mu}$) is always a Killing
  vector which can be timelike or null. These two cases are treated
  separately.
  
\item[III] One can get an expression of all the gauge field strengths of the
  theory using the Killing equation for those scalar bilinears: $\Omega_{\mu}$
  is usually of the form $F_{\mu\nu}V^{\nu}$ and, then
  Eq.~(\ref{eq:Killingeq}) tells us that $F_{\mu\nu}V^{\nu}\sim \nabla_{\mu}
  \log M$.  When $V$ is timelike this determines completely $F$ and, when it
  is null, it determines the general form of $F$. Of course,
  Eq.~(\ref{eq:Killingeq}) is an oversimplified KSE and in real-life
  situations there are additional scalar factors, $SU(N)$ indices etc.
  
\item[IV] The Maxwell and Einstein equations and Bianchi identities are
  imposed on those field strengths $F$, getting second order equations for the
  scalar bilinears $M$. 
  
\item[V] The KSIs guarantee that these three different sets of equations (plus
  the equations of the scalar fields, if any) are complicated combinations a a
  reduced number of simple equations involving a reduced number of scalar
  unknowns. Solving these equations for the scalar unknowns gives full
  solutions of the theory. The tricky part is, usually, identifying the
  right variables that satisfy simple equations and finding these equations as
  combinations of the Maxwell, Einstein etc. equations.

\item[VI] Finally, with the results obtained, the KSEs have to be solved,
  which may lead to additional conditions on the fields.

\end{itemize}

Let us see how this recipe works in the examples considered before.

%%%%%%%%%%%%%%%%%%%%%%%%%%%%%%%%%%%%%%%%%%%%%%%%%%%%%%%%%%%%%%%%%%%%%%
%%%%%%%%%%%%%%%%%%%%%%%%%%%%%%%%%%%%%%%%%%%%%%%%%%%%%%%%%%%%%%%%%%%%%%
%%%%%%%%%%%%%%%%%%%%%%%%%%%%%%%%%%%%%%%%%%%%%%%%%%%%%%%%%%%%%%%%%%%%%%

\subsection{Example: $N=1,d=4$ Supergravity.}

With one ({Majorana}) {Killing spinor $\epsilon$} the only bilinear that one
can construct is a real vector bilinear $V_{\mu}$ which is always null.
$V_{\mu}$ is also covariantly constant (i.e.~it is a Killing vector and
$V_{\mu}dx^{\mu}$ is an exact 1-form, which allows us to write
$V_{\mu}dx^{\mu}=du$):

\begin{equation}
\delta_{{\epsilon}}{\psi_{\mu}}=
\nabla_{\mu}{\epsilon}=0\, ,\,\,\, \Rightarrow   
\nabla_{\mu} {V_{\nu}}=0\, ,\,\,\, 
{R^{\mu}{}_{\nu}}{V^{\nu}}=0\,
,\,\,\,\, ({\bar{\epsilon}}
{R^{\mu}{}_{a}}\gamma^{a}{\epsilon}=0)\, .
\end{equation}

\noindent
All the metrics with covariantly constant null vectors are Brinkmann pp-waves
and have the form

\begin{equation}
ds^{2} = 2 du (dv + K du +A_{\underline{i}}  d x^{i})
+\tilde{g}_{\underline{i}\underline{j}}  dx^{i} d x^{j} \, ,
\end{equation}

\noindent
where all the components are independent of $v$, where $v$ is defined by
$V^{\mu} \partial_{\mu}\equiv \partial/\partial v$.

It can be checked that for all these metrics the KSE has solutions. These,
then, are all the supersymmetric field configurations of $N=1,d=4$ SUGRA, but
only those with $R_{\mu\nu}=0$ are supersymmetric solutions.

%%%%%%%%%%%%%%%%%%%%%%%%%%%%%%%%%%%%%%%%%%%%%%%%%%%%%%%%%%%%%%%%%%%%%%
%%%%%%%%%%%%%%%%%%%%%%%%%%%%%%%%%%%%%%%%%%%%%%%%%%%%%%%%%%%%%%%%%%%%%%
%%%%%%%%%%%%%%%%%%%%%%%%%%%%%%%%%%%%%%%%%%%%%%%%%%%%%%%%%%%%%%%%%%%%%%

\subsection{Example: $N=2,d=4$ Supergravity.}

With two Weyl spinors\footnote{In this theory one can use pairs of Majorana or
  Weyl spinors or single Dirac spinors. We now use, for convenience, pairs of
  Weyl spinors.} $\epsilon^{I}$ one can construct the following independent
bilinears

\begin{itemize}
\item A complex scalar $\bar{\epsilon}^{I}\epsilon^{J}\equiv
  M\varepsilon^{IJ}$
\item A Hermitean matrix of null vectors $V^{I}{}_{J\, \mu}\equiv
  i\bar{\epsilon}^{I} \gamma_{\mu}\epsilon_{J}$
\end{itemize}

The KSEs imply the following equations for the bilinears:

\begin{eqnarray}
\nabla_{\mu}M & \sim &  
F^{+}{}_{\mu\nu}V^{I}{}_{I}{}^{\nu}\, ,\label{eq:FV}\\
& & \nonumber \\
\nabla_{\mu}V^{I}{}_{J\, \nu} & \sim &
\delta^{I}{}_{J}[M F^{+}{}_{\mu\nu} 
+M^{*} F^{-}{}_{\mu\nu}]
-\Phi_{KJ\, (\mu}{}^{\rho}
\varepsilon^{KI}F^{-}{}_{\nu)\rho}
-\Phi^{IK}{}_{(\mu|}{}^{\rho}
 \varepsilon_{KJ} F^{+}{}_{|\nu)\rho}\ ,\\
\end{eqnarray}

\noindent
so the vector $V^{\mu}\equiv V^{I}{}_{I}{}^{\mu}$ is Killing and the other
three are exact forms.  The Fierz identities tell us that $V^{\mu}V_{\mu}\sim
|M|^{2}\geq 0$ can be timelike or null. When it is timelike,
$V^{\mu}\partial_{\mu}\equiv \sqrt{2} \partial/\partial t$  and the metric can
be put in the \textit{conformastationary} form 

\begin{equation}
\label{eq:conformastat}
ds^{2}  =  |M|^{2} (dt+\omega)^{2} -|M|^{-2} d\vec{x}^{\, 2}\, , 
\end{equation}

\noindent
where, for consistency, the 1-form $\omega$ has to be related to $M$ by

\begin{equation}
\label{eq:doeq}
d\omega= i|{M}|^{-2}{}^{\star} [{{M}  d{M}^{*}-{\rm c.c.}}]\, .
\end{equation}

\noindent
On the other hand, Eq.~(\ref{eq:FV}) gives

\begin{equation}
F^{+}  \sim  |M|^{-2} \{V\wedge dM +i {}^{\star}[V\wedge dM] \}\, .
\end{equation}

The KSIs are satisfied if Eq.~(\ref{eq:doeq}) is satisfied. It can be seen
that, then, any metric and 2-form field strength of the above form admit
Killing spinors. On the other hand, all the equations of motion are
combinations of the simple equation in 3-dimensional Euclidean space

\begin{equation}
\vec{\nabla}^{\,2}{M}^{-1}=0\, .  
\end{equation}

\noindent
Thus, solving this equation for some $M$ gives us a supersymmetric solution of
all the equations of motion (all the fields are determined by $M$). These
solutions of the Einstein-Maxwell theory are the Israel-Wilson-Perj\'es family
\cite{kn:IW,kn:Pe}.

The case in which $V$ is null is very similar to the $N=1$ case and we will
not study it here in detail for lack of space.

%%%%%%%%%%%%%%%%%%%%%%%%%%%%%%%%%%%%%%%%%%%%%%%%%%%%%%%%%%%%%%%%%%%%%%
%%%%%%%%%%%%%%%%%%%%%%%%%%%%%%%%%%%%%%%%%%%%%%%%%%%%%%%%%%%%%%%%%%%%%%
%%%%%%%%%%%%%%%%%%%%%%%%%%%%%%%%%%%%%%%%%%%%%%%%%%%%%%%%%%%%%%%%%%%%%%

\section{Tod's problem in $N=4,d=4$ supergravity}

\begin{figure}[htbp]
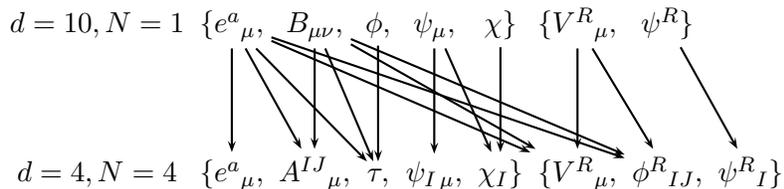

     \centering
     \setlength{\abovecaptionskip}{0mm}%
$
 \begin{psmatrix}[rowsep=1.5,colsep=0.2]
 d=10, N=1
 &
 [name=e10]{\{e^{a}{}_{\mu},}
 &
 [name=B10]{B_{\mu\nu},}
 &
 [name=d10]{\phi,}
 &
 [name=p10]{\psi_{\mu},}
 &
 [name=c10]{\chi\}}
 &
 [name=V10]{\{V^{R}{}_{\mu},}
 & 
 [name=pm10]{\psi^{R}\}}
 & 
 &
 \\
 d=4, N=4 
 &
 [name=e4]{\{e^{a}{}_{\mu},}
 &
 [name=A4]{A^{IJ}{}_{\mu},}
 &
 [name=t4]{\tau,}
 &
 [name=p4]{\psi_{I\, \mu},}
 & 
 [name=c4]{\chi_{I}\}}
 & 
 [name=V4]{\{V^{R}{}_{\mu},}
 &
 [name=f4]{\phi^{R}{}_{IJ},}
 & 
 [name=pm4]{\psi^{R}{}_{I}\}}
 &
% Links
   \everypsbox{\scriptstyle}
   \psset{arrows=->,nodesep=0.1}
 \ncline{e10}{e4}^[tpos=0.3]{}%
 \ncline{e10}{V4}^[tpos=0.3]{}%
 \ncline{e10}{A4}^[tpos=0.3]{}%
 \ncline{e10}{t4}^[tpos=0.3]{}%
 \ncline{e10}{f4}^[tpos=0.3]{}%
 \ncline{B10}{A4}^[tpos=0.3]{}%
 \ncline{B10}{V4}^[tpos=0.3]{}%
 \ncline{B10}{t4}^[tpos=0.3]{}%
 \ncline{B10}{f4}^[tpos=0.3]{}%
 \ncline{d10}{t4}^[tpos=0.3]{}%
 \ncline{p10}{p4}^[tpos=0.3]{}%
 \ncline{p10}{c4}^[tpos=0.3]{}%
 \ncline{c10}{c4}^[tpos=0.3]{}%
 \ncline{V10}{V4}^[tpos=0.3]{}%
 \ncline{V10}{f4}^[tpos=0.3]{}%
 \ncline{pm10}{pm4}^[tpos=0.3]{}%
 \end{psmatrix}
$
 \caption{Relation between the fields of $N=1,d=10$ SUGRA $N=4,d=4$ SUGRA. The
   fields in curly brackets belong to the same supermultiplet. Both in $d=10$
   and $d=4$ there a supergravity multiplet containing the graviton and vector
   supermultiplets, but the 4-dimensional vector supermultiplets originate from
   both the $d=10$ supergravity and vector supermultiplets. The $I,J=1,\cdots,
   4$ indices are $SU(4)$ indices related to the six internal dimensions using
   the isomorphism between $SO(6)$ and $SU(4)$. The $R,S=1,\cdots,22$ indices
   count the vector supermultiplets: 6 of them coming from the supergravity
   multiplet and 16 from 10-dimensional vector supermultiplets. }
 \label{fig:fields}
   \end{figure}

This theory can be obtained by toroidal compactification on $T^{6}$ of
$N=1,d=10$ SUGRA \cite{Chamseddine:1980cp} (the effective field theory of the
Heterotic String) and subsequent (consistent) truncation of the matter vector
fields. The 10- and 4-dimensional fields are related as indicated in
Fig.~\ref{fig:fields}.

A special role is played by the \textit{axidilaton} field $\tau=a+ie^{-\phi}$,
where $a$ is dual to the 4-dimensional Kalb-Ramond 2-form and plays the role
of local $\theta$ parameter and $\phi$ is the 4-dimensional dilaton, which
plays its usual role of local coupling constant. 
  
It is convenient to start by studying the {\it pure} supergravity theory
(without the vector supermultiplets) \cite{Cremmer:1977tt}, for simplicity.
The theory has global $SU(4)$ symmetry (duality) and, furthermore, only at the
level of the equations of motion, an $SL(2,\mathbb{R})$ invariance (S~duality)
that rotates Maxwell equations into Bianchi identities and acts on the
axidilaton according to

\begin{equation}
\tau^{\prime} =
{\displaystyle\frac{\alpha\tau +\beta}{\gamma \tau+\delta}}\, , 
\hspace{1cm}
\alpha\delta-\gamma\beta=1\, .
\end{equation}

\noindent
Observe that the $N=2$ and $N=1$ are included as truncations.
  
The bosonic action of the theory is

\begin{equation}
  \label{eq:N4d4SUGRAaction}
S=\int d^{4}x\sqrt{|g|}\left[R 
+{\textstyle\frac{1}{2}}
\frac{\partial_{\mu}\tau\,\partial^{\mu}\tau^{*}}{(\Im {\rm m}\,\tau)^{2}}
-{\textstyle\frac{1}{16}} \Im {\rm m}\,\tau F^{IJ\, \mu\nu} F_{IJ\, \mu\nu}
-{\textstyle\frac{1}{16}} \Re {\rm e}\, \tau
F^{IJ\, \mu\nu} {}^{\star}\!\!F_{IJ\, \mu\nu}
\right]\, .  
\end{equation}

It is convenient to denote the equations of motion by

\begin{equation}
\mathcal{E}_{a}{}^{\mu}\equiv 
-\frac{1}{2\sqrt{|g|}}\frac{\delta S}{\delta e^{a}{}_{\mu}}\, ,
\hspace{1cm}
\mathcal{E}\equiv -\frac{2\Im {\rm m}\tau}{\sqrt{|g|}}
\frac{\delta S}{\delta \tau}\, ,
\hspace{1cm}
\mathcal{E}^{IJ\, \mu}\equiv 
\frac{8}{\sqrt{|g|}}\frac{\delta S}{\delta A_{IJ\, \mu}}\, .
\end{equation}

The Maxwell equation $\mathcal{E}^{IJ\, \mu}$ transforms as an
$SL(2,\mathbb{R})$ doublet together with the Bianchi identity 

\begin{equation}
\label{eq:B}
\mathcal{B}^{IJ\, \mu}\equiv \nabla_{\nu}{}^{\star}F^{IJ\, \nu\mu}\, .
\end{equation}

For vanishing fermions, the supersymmetry transformation rules of the
gravitini and dilatini, generated by 4 spinors $\epsilon_{I}$ of
negative chirality, are

\begin{eqnarray}
\label{eq:fermionsusyvariations}
\delta_{\epsilon} \psi_{I\, \mu} & = &
\mathcal{D}_{\mu}\epsilon_{I} 
-{\textstyle\frac{i}{2\sqrt{2}}}  
(\Im {\rm m}\,\tau)^{1/2}F_{IJ}{}^{+}{}_{\mu\nu}\gamma^{\nu}\epsilon^{J}\, ,\\
& & \nonumber \\
\delta_{\epsilon} \chi_{I} & = & 
{\textstyle\frac{1}{2\sqrt{2}}}\frac{\not\!\partial\tau}{\Im {\rm m}\,\tau} 
\epsilon_{I}
-{\textstyle\frac{1}{8}} 
(\Im {\rm m}\,\tau)^{1/2}\!\not\!F_{IJ}{}^{-}\epsilon^{J}\, ,
\end{eqnarray}

\noindent
where $\mathcal{D}$ is the Lorentz plus $U(1)$ covariant derivative and where
the $U(1)$ connection is given by 

\begin{equation}
\label{eq:Q}
Q_{\mu}\equiv {\textstyle\frac{1}{4}}
\frac{\partial_{\mu} \Re {\rm e}\, \tau}{\Im {\rm m}\,\tau}\, .
\end{equation}

The supersymmetry transformation rules of the bosonic fields take the form

\begin{eqnarray}
\delta_{\epsilon} e^{a}{}_{\mu} & = & 
-{\textstyle\frac{i}{4}}( \bar{\epsilon}^{I} \gamma^{a} \psi_{I\, \mu}
+\bar{\epsilon}_{I} \gamma^{a} \psi^{I}{}_{\mu})\, ,
\label{eq:susytranseam}\\ 
& & \nonumber \\
\delta_{\epsilon} \tau  & = & 
-{\textstyle\frac{i}{\sqrt{2}}}
 \Im {\rm m} \tau \bar{\epsilon}^{I} \chi_{I}\, , 
\label{eq:susytranstau}\\ 
& & \nonumber \\
\delta_{\epsilon} A_{IJ\, \mu} & = & 
\frac{\sqrt{2}}{(\Im {\rm m} \tau)^{1/2}} 
\left[
\bar{\epsilon}_{[I} \psi_{J]\, \mu} 
+{\textstyle\frac{i}{\sqrt{2}}} \bar{\epsilon}_{[I}\gamma_{\mu} \chi_{J]} 
+{\textstyle\frac{1}{2}} \epsilon_{IJKL} 
\left(\bar{\epsilon}^{K} \psi^{L}{}_{\mu} + 
{\textstyle{\frac{i}{\sqrt{2}}}} \bar{\epsilon}^{K}
\gamma_{\mu} \chi^{L} \right) 
\right]\, .
\label{eq:susytransaij}
\end{eqnarray}

Given $N$ chiral commuting spinors $\epsilon_{I}$ and their complex conjugates
$\epsilon^{I}$ we can constructed the following independent bilinears:

\begin{enumerate}
\item A complex, antisymmetric, matrix of scalars

\begin{equation}
M_{IJ}\equiv \bar{\epsilon}_{I}\epsilon_{J}\, ,  
\hspace{1cm}
M^{IJ}\equiv \bar{\epsilon}^{I}\epsilon^{J}=(M_{IJ})^{*}\, ,
\end{equation}

\item A complex matrix of vectors

\begin{equation}
V^{I}{}_{J\, a}\equiv i\bar{\epsilon}^{I}\gamma_{a}\epsilon_{J}\, ,  
\hspace{1cm}
V_{I}{}^{J}{}_{a}\equiv i\bar{\epsilon}_{I}\gamma_{a}\epsilon^{J}
=(V^{I}{}_{J\, a})^{*}\, ,
\end{equation}

\noindent
which is Hermitean:

\begin{equation}
(V^{I}{}_{J\, a})^{*}=V_{I}{}^{J}{}_{a} = V^{J}{}_{I\, a}
=(V^{I}{}_{J\, a})^{T}\, .  
\end{equation}

\end{enumerate}

Using the supersymmetry transformation rules of the bosonic fields, one can
find the KSIs of this theory, associated to the gravitini and dilatini,
respectively. However, just as in the $N=2,d=4$ example, since the Bianchi
identities do not appear in these equations, they break S-duality covariance.
This covariance can be restored by hand or re-deriving the KSIs from the KSEs
integrability conditions. The result is

\begin{eqnarray}
\label{eq:ks2-2}
\mathcal{E}^{\mu}{}_{a}\gamma^{a}\epsilon_{I}    
-\frac{i}{\sqrt{2}(\Im {\rm m}\, \tau)^{1/2}}
(\mathcal{E}_{IJ}{}^{\mu}  
-\tau^{*}\mathcal{B}_{IJ}{}^{\mu})\epsilon^{J} & = & 0\, ,\\
& & \nonumber \\
\label{eq:ks2-3}
\mathcal{E}^{*}\epsilon_{I}  
-\frac{1}{\sqrt{2}(\Im {\rm m}\, \tau)^{1/2}}
(\not\!\!\mathcal{E}_{IJ} -\tau\not\!\!\mathcal{B}_{IJ})\epsilon^{J} 
& = & 0\, .
\end{eqnarray}

It is useful to derive tensorial equations from these KSIs. Combining them we
arrive to the following, which are chosen among the many possible tensorial
KSIs by their interest. For timelike $V^{a}\equiv V^{I}{}_{I}$ we get

\begin{eqnarray}
\mathcal{E}^{ab} 
-{\textstyle\frac{1}{2}}\Im {\rm m}\, \mathcal{E} V^{a}V^{b} 
-\frac{1}{\sqrt{2}} (\Im {\rm m}\, \tau)^{1/2}
\Im {\rm m}\, (M^{IJ}\mathcal{B}_{IJ}{}^{a})V^{b} & = & 0\, , 
\label{eq:ksi3} \\
& & \nonumber \\
\mathcal{E}^{*} V^{a} -\frac{i}{\sqrt{2}(\Im {\rm m}\, \tau)^{1/2}}
M^{IJ}(\mathcal{E}_{IJ}{}^{a} -\tau\mathcal{B}_{IJ}{}^{a}) 
& = & 0\, ,
\label{eq:ksi4} \\
& & \nonumber \\
\Im {\rm m} [M^{IJ}(\mathcal{E}_{IJ}{}^{a} 
-\tau^{*}\mathcal{B}_{IJ}{}^{a})] & = & 0\, .
\label{eq:ksi5}
\end{eqnarray}

\noindent
Observe that the first equation implies the off-shell vanishing of all the
Einstein equations with one or two spacelike components. Further, the Einstein
equation is automatically satisfied when the Maxwell, Bianchi and complex
scalar equations are satisfied and the scalar equation is automatically
satisfied when the Maxwell and Bianchi are.

When $V^{a}$ is null (we denote it by $l^{a}$), all the spinors $\epsilon_{I}$
are proportional and we can parametrize all of them by
$\epsilon_{I}=\phi_{I}\epsilon$, where $\phi^{I}\phi_{I}=1$. In order to
construct tensor bilinears we define an auxiliary spinor $\eta$ normalized by
$ \bar{\epsilon}\eta=\frac{1}{2}$. With these two spinors we can construct a
standard complex null tetrad

\begin{equation}
\label{eq:nulltetraddef}
l_{\mu}=i\bar{\epsilon^{*}}\gamma_{\mu}\epsilon\, ,
\hspace{.5cm}
n_{\mu}=i\bar{\eta^{*}}\gamma_{\mu}\eta\, ,
\hspace{.5cm}
m_{\mu}=i\bar{\epsilon^{*}}\gamma_{\mu}\eta=
i\bar{\eta}\gamma_{\mu}\epsilon^{*}\, ,
\hspace{.5cm}
m_{\mu}^{*}=i\bar{\epsilon}\gamma_{\mu}\eta^{*}=
i\bar{\eta^{*}}\gamma_{\mu}\epsilon\, .
\end{equation}

Then, in the null case, the KSIs take the form

\begin{eqnarray}
\label{eq:ksi6}
(\mathcal{E}^{\mu}{}_{a}
-{\textstyle\frac{1}{2}}e_{a}{}^{\mu}\mathcal{E}^{\rho}{}_{\rho})\, l^{a} = 
(\mathcal{E}^{\mu}{}_{a}
-{\textstyle\frac{1}{2}}e_{a}{}^{\mu}\mathcal{E}^{\rho}{}_{\rho})\, m^{a} 
& = & 0\, ,\\
& & \nonumber \\
\label{eq:ksi7}
\mathcal{E} & = & 0\, ,\\
& & \nonumber \\
\label{eq:ksi8}
(\mathcal{E}_{IJ}{}^{\mu} 
-\tau^{*}\mathcal{B}_{IJ}{}^{\mu})\phi^{J} & = & 0\, .
\end{eqnarray}

\noindent
In this case supersymmetry implies that the scalar equations of motion are
automatically satisfied.  We are not going to work out here the null case,
since it was treated completely in Ref.~\cite{Tod:1995jf}.

We are now ready to follow the recipe to find all the supersymmetric
configurations of this theory. The first step consists in finding (Killing)
equations for the spinor bilinears. From the vanishing of the gravitini
supersymmetry transformation rule we find

\begin{eqnarray}
\mathcal{D}_{\mu}M_{IJ} & = & 
{\textstyle\frac{1}{\sqrt{2}}}(\Im {\rm m}\,\tau)^{1/2}  
F_{K[I|}{}^{+}{}_{\mu\nu} V^{K}{}_{|J]}{}^{\nu}\, , \label{eq:dm}\\
& & \nonumber \\
\mathcal{D}_{\mu} V^{I}{}_{J\, \nu} & = & 
-{\textstyle\frac{1}{2\sqrt{2}}}(\Im {\rm m}\,\tau)^{1/2}
\left[M_{KJ}F^{KI\, -}{}_{\mu\nu} +M^{IK}F_{JK}{}^{+}{}_{\mu\nu}
\right. \nonumber\\
& & \nonumber \\
& & \left.
-\Phi_{KJ\, (\mu}{}^{\rho}F^{KI\, -}{}_{\nu)\rho}
-\Phi^{IK}{}_{(\mu|}{}^{\rho}F_{KI}{}^{+}{}_{|\nu)\rho}
\right]\, , \label{eq:dv} 
\end{eqnarray}

\noindent
and from that of the dilatini, we find

\begin{eqnarray}
\label{eq:vt}
V^{K}{}_{I}\cdot\partial\tau -
{\textstyle\frac{i}{2\sqrt{2}}} 
(\Im {\rm m}\,\tau)^{3/2}F_{IJ}{}^{-}\cdot \Phi^{KJ} & = & 0\, , \\
& & \nonumber \\
\label{eq:mt}
F_{IJ}{}^{-}{}_{\rho\sigma}V^{J}{}_{K}{}^{\sigma }
+{\textstyle\frac{i}{\sqrt{2}}} (\Im {\rm m}\,\tau)^{-3/2} 
\left(M_{IK}\partial_{\rho}\tau -
\Phi_{IK\, \rho}{}^{\mu}\partial_{\mu}\tau\right) & = & 0\, . 
\end{eqnarray}

It is immediate to see that $V\equiv V^{I}{}_{I}$ is a Killing vector and 
that 

\begin{equation}
\label{eq:vt0}
V^{\mu}\partial_{\mu} \tau =0\, .  
\end{equation}

Further, using Eq.~(\ref{eq:dm}) and the antisymmetric part of
Eq.~(\ref{eq:mt}) we find 

\begin{equation}
\label{eq:FSRV}
F_{SR}{}^{-}{}_{\mu\nu}V^{\nu} =
-\frac{\sqrt{2}i}{(\Im {\rm m}\,\tau)^{3/2}}M_{SR}\partial_{\mu}\tau
-\frac{\sqrt{2}}{(\Im {\rm m}\,\tau)^{1/2}}\varepsilon_{SRIJ}
\mathcal{D}_{\mu}M^{IJ}\, ,
\end{equation}

\noindent
which determines completely the vector field strengths in terms of the scalar
bilinears, $\tau$ and the Killing vector $V^{a}$ when this is timelike. In the
null case, this equation gives us important constraints on the form of the
field strengths, but does not completely determine them. From now on we will
focus on the timelike case since it illustrates our procedure best. In this
case we can write the metric in the conformastationary form
Eq.~(\ref{eq:conformastat}), but, while in the $N=2,d=4$ case one could show
that three of the vector bilinears where exact 1-forms and then the metric on
the constant-time slices could be chosen to be Euclidean, in the $N=4,d-4$ case
this is not possible and we have to live with a non-trivial 3-dimensional
metric $\gamma_{\underline{i}\underline{j}}$. Thus

\begin{equation}
\label{eq:metric}
ds^{2} = |M|^{2}(dt+\omega)^{2} 
-|M|^{-2}\gamma_{\underline{i}\underline{j}}dx^{i}dx^{j}\, ,
\hspace{1cm}
i,j=1,2,3\, ,
\end{equation}

\noindent
where $\omega$ has to satisfy the equation

\begin{equation}
\label{eq:omega}
d\omega = {\textstyle\frac{1}{\sqrt{2}}}\Omega =
 {\textstyle\frac{i}{2\sqrt{2}}}|M|^{-4}\,
 {}^{\star}\!\!\left[(M^{IJ}\mathcal{D}M_{IJ}
 -M_{IJ}\mathcal{D}M^{IJ})\wedge \hat{V}\right]\, .
% = \sqrt{2}|M|^{-2}\, {}^{\star}\!\left[(Q -\xi)\wedge V\right]\, .  
\end{equation}

Having the field strengths expressed in terms of the scalars $M^{IJ},\tau$, we
move on to the next step and impose the Maxwell equations and Bianchi
identities on them, to obtain equations that only involve those scalars. We
also substitute the field strengths into the $\tau$ equation, obtaining
another equation that only involves $M^{IJ}$ and $\tau$. Now comes the magic
of supersymmetry: these three sets of equations are combinations of just two
sets of much simpler equations in the 3-dimensional metric
$\gamma_{\underline{i}\underline{j}}$:

\begin{eqnarray}
n^{IJ}_{(3)} & \equiv & (\nabla_{\underline{i}} +4i\xi_{\underline{i}}) 
\left(\frac{\partial^{\underline{i}}N^{IJ}}{|N|^{2}} \right)\, ,
\label{eq:nequation3}\\
& & \nonumber \\
e^{*}_{(3)} & \equiv & (\nabla_{\underline{i}} +4i\xi_{\underline{i}}) 
\left(\frac{\partial^{\underline{i}}\tau}{|N|^{2}} \right)\, ,
\label{eq:eequation3}
\end{eqnarray}

\noindent
where $N^{IJ}\equiv (\Im{\rm m}\tau)^{1/2}M^{IJ}$ and $\xi$ is defined by

\begin{equation}
\xi \equiv 
\textstyle{\frac{i}{4}}|M|^{-2}(M_{IJ}dM^{IJ}-M^{IJ}dM_{IJ})\, ,
\label{eq:xidef} 
\end{equation}

\noindent 
and acts as a $U(1)$ connection.

In fact, we can write all the components of the equations of motion define
above in terms of these two

\begin{eqnarray} 
\mathcal{E}_{00} & = & 
|M|^{2} \left[|M|^{2} \Im {\rm m}e^{*}_{(3)} 
-2\Re {\rm e}\, (N_{KL}n^{KL}_{(3)})
+{\textstyle\frac{1}{2}}e_{k}{}^{k}\right]\, ,
\label{eq:einsteinequation003}\\
& & \nonumber \\
\mathcal{E}_{0i} & = & 0\, ,
\label{eq:einsteinequation0i3}\\
& & \nonumber \\
\mathcal{E}_{ij} & = & |M|^{2}(e_{ij}
-{\textstyle\frac{1}{2}}\delta_{ij}e_{k}{}^{k})\, ,
\label{eq:einsteinequationij3}\\
\mathcal{B}^{IJ\, a} & = & -\sqrt{2}|M|^{2}V^{a}
\left\{
\frac{N^{IJ}+\tilde{N}^{IJ}}{\Im {\rm m}\tau}\Re {\rm e}\, e_{(3)}
-i(n^{IJ}_{(3)}-\tilde{n}^{IJ}_{(3)})
\right\}\, ,
\label{eq:bianchiequation3} \\
& & \nonumber \\
\mathcal{E}^{IJ\, a} & = & -\sqrt{2}|M|^{2}V^{a}
\left\{
\frac{N^{IJ}+\tilde{N}^{IJ}}{\Im {\rm m}\tau}\Re {\rm e}\, 
(\tau e_{(3)})
-i(\tau^{*}n^{IJ}_{(3)}-\tau\tilde{n}^{IJ}_{(3)})
\right\}\, .
\label{eq:maxwellequation3} \\
& & \nonumber \\
\mathcal{E} & = & -|M|^{2} \left[|M|^{2} e_{(3)} 
+2i N_{KL}\tilde{n}^{KL}_{(3)}\right]\, , 
\label{eq:tauequation3} 
\end{eqnarray}

\noindent
and a set of equations $e_{ij}$ defined by

\begin{equation}
e_{ij} \equiv R_{ij}(\gamma) -2\partial_{(i}\left(\frac{N^{IJ}}{|N|}\right)
\partial_{j)}\left(\frac{N_{KL}}{|N|}\right)
(\delta^{KL}{}_{IJ} -\mathcal{J}^{K}{}_{I}\mathcal{J}^{L}{}_{J})\, ,  
\end{equation}

\noindent
and which have to vanish in order to satisfy the KSIs and have
supersymmetry\footnote{The integrability condition of the equation for
  $\omega$ has to be satisfied as well in order to have supersymmetry. WE are
  going to discuss it later.}.  These equations are conditions for the
3-dimensional metric $\gamma_{\underline{i}\underline{j}}$, but are not easy
to solve directly.  We have to substitute our results into the original KSEs
or into their integrability conditions. The solution one finds is that, in
order to solve the $e_{ij}=0$ equations have supersymmetry, the 3-dimensional
metric has to take the form

\begin{equation}
\label{eq:metric2}
\gamma_{\underline{i}\underline{j}}dx^{\underline{i}}dx^{\underline{j}}
= dx^{2}+2e^{2U(z,z^{*})}dzdz^{*}\, ,
\end{equation}

\noindent
and the connection $\xi$ has to take the form

\begin{equation}
\label{eq:xiU}
\xi = \pm{\textstyle\frac{i}{2}}
(\partial_{\underline{z}}Udz -\partial_{\underline{z}^{*}}Udz^{*}) 
+{\textstyle\frac{1}{2}}d\lambda  (x,z,z^{*})\, .  
\end{equation}

\noindent
Since $\xi$ is defined in terms of the  $M^{IJ}$ scalars, this is a condition
that these scalars have to fulfill, on top of Eqs.~(\ref{eq:nequation3},\ref{eq:eequation3}).

Further, to have supersymmetry, the integrability condition for the equation
defining $\omega$ has to be satisfied as well. It takes the form

\begin{equation}
\label{eq:susyequation2bis}
\nabla_{\underline{i}}\left(\frac{Q^{\underline{i}}
-\xi^{\underline{i}}}{|M|^{2}}\right)=0\, .  
\end{equation}

The timelike case now has been completely solved. Let us put together the
results: any supersymmetric configuration of $N=4,d=4$ supergravity in this
class is given by a set of 7 complex functions $M^{IJ},\tau$ which have to
satisfy the following conditions:

\begin{enumerate}
\item $M^{[IJ}M^{K]L}=0$. This is a condition that the scalar bilinears
  satisfy due to the Fierz identities.
\item $|M|^{2}\neq 0$. We have assumed this, as definition of the timelike
  case ($V^{2}\sim |M|^{2}>0$).
\item Eq.~(\ref{eq:susyequation2bis}) has to be satisfied.
\item $\xi$ has to take the form Eq.~(\ref{eq:xiU}).
\end{enumerate}

Given 7 complex functions satisfying these conditions, then, a supersymmetric
field configuration of $N=4,d=4$ is given by the metric
Eqs.~(\ref{eq:metric},\ref{eq:metric2}) and the field strengths
Eq.~(\ref{eq:FSRV}). These field configurations will be supersymmetric
solutions if the expressions Eqs.~(\ref{eq:nequation3},\ref{eq:eequation3})
vanish.

This is the main result in the timelike case. 

Now comes the problem of finding sets of 7 complex functions satisfying the
above conditions, which is not an easy. We have been able to find two families
of supersymmetric solutions based on the \textit{Ansatz} for the $M^{IJ}$s

\begin{equation}
\label{eq:ansatzm}
M_{IJ}= e^{i\lambda (x,z,z^{*})} M (x,z,z^{*}) k_{IJ}(z)\, ,  
\hspace{.5cm}
M=M^{*}\, ,
\hspace{.5cm}
\lambda=\lambda^{*}\, ,
\hspace{.5cm}
k_{[IJ}k_{K]L}=0\, .
\end{equation}

\noindent
which give a connection $\xi$ of the form  Eq.~(\ref{eq:xiU}) with 

\begin{equation}
U= +\ln{|k|}\, ,
\hspace{1cm}
|k|^{2}\equiv k^{IJ}(z^{*})k_{IJ}(z)\, . 
\end{equation}

This \textit{Ansatz} satisfies all the conditions except for
Eq.~(\ref{eq:susyequation2bis}). In the following two cases, at least, this
last condition is also satisfied:

\begin{enumerate}
\item If the $k_{IJ}$ are constants, then, normalizing $|k|^{2}=1$ for
  simplicity, $\xi=\frac{1}{2}d\lambda$ and $U=0$.  This case was considered
  by Tod in Ref.~\cite{Tod:1995jf} and studied in detail in
  Ref.~\cite{Bergshoeff:1996gg}. Defining $ \mathcal{H}_{1}\equiv [(\Im {\rm
    m}\, \tau)^{1/2} e^{-i\lambda}M]^{-1}$, and $\tau= \mathcal{H}_{1}/
  \mathcal{H}_{2}$ we get solutions if
  $\partial_{\underline{i}}\partial_{\underline{i}} \mathcal{H}_{1}=
  \partial_{\underline{i}}\partial_{\underline{i}} \mathcal{H}_{2}=0$.
  
\item With $e^{i\lambda}=M=1$ and constant $\tau$ we solve all constraints and
  all equations using the holomorphicity of the $k_{IJ}$s. The metric takes
  the form

\begin{equation}
ds^{2}=|k|^{2}(dt +\omega)^{2} -|k|^{-2}dx^{2} -2dzdz^{*}\, .    
\end{equation}

The metric and the supersymmetry projectors correspond to stationary strings
lying along the coordinate $x$, in spite of the trivial axion field $\Re{\rm
  e}\tau$. These solutions clearly deserve more study. Observe that this
family is precisely the one that cannot be embedded in $N=2,d=4$ supergravity
plus matter fields \cite{Ferrara:1995ih} and it is genuinely $N=4$.

\end{enumerate}

%%%%%%%%%%%%%%%%%%%%%%%%%%%%%%%%%%%%%%%%%%%%%%%%%%%%%%%%%%%%%%%%%%%%%%
%%%%%%%%%%%%%%%%%%%%%%%%%%%%%%%%%%%%%%%%%%%%%%%%%%%%%%%%%%%%%%%%%%%%%%
%%%%%%%%%%%%%%%%%%%%%%%%%%%%%%%%%%%%%%%%%%%%%%%%%%%%%%%%%%%%%%%%%%%%%%

\section{Conclusions}

The landscape approach offers an interesting, even if controversial, point of
view over the vacuum selection problem. It also gives additional reasons to
work on the problem of classification of supersymmetric solutions, whose
4-dimensional structure we have reviewed in this talk, emphasizing the
difference between general supersymmetric configurations and solutions and
showing how the KSIs can be used in this problem. We have applied the recipes
to an interesting case: pure $N=4,d=4$ supergravity, but is should be clear
that the same procedure could be used in more general contexts ($N=4,d=4$
coupled to matter, gauged etc. and other 4-dimensional theories
\cite{kn:BHO,kn:MO}). We also expect some of the techniques could also be of
use in solving the much more complicated 11- and 10-dimensional problems
\cite{Gauntlett:2002fz,MacConamhna:2005vt}.

%%%%%%%%%%%%%%%%%%%%%%%%%%%%%%%%%%%%%%%%%%%%%%%%%%%%%%%%%%%%%%%%%%%%%%
%%%%%%%%%%%%%%%%%%%%%%%%%%%%%%%%%%%%%%%%%%%%%%%%%%%%%%%%%%%%%%%%%%%%%%
%%%%%%%%%%%%%%%%%%%%%%%%%%%%%%%%%%%%%%%%%%%%%%%%%%%%%%%%%%%%%%%%%%%%%%
%%%%%%%%%%%%%%%%%%%%%%%%%%%%%%%%%%%%%%%%%%%%%%%%%%%%%%%%%%%%%%%%%%%%%%
%%%%%%%%%%%%%%%%%%%%%%%%%%%%%%%%%%%%%%%%%%%%%%%%%%%%%%%%%%%%%%%%%%%%%%

\section*{Acknowledgments}

This work has been supported in part by the Spanish grant BFM2003-01090.

%%%%%%%%%%%%%%%%%%%%%%%%%%%%%%%%%%%%%%%%%%%%%%%%%%%%%%%%%%%%%%%%%%%%%%
%%%%%%%%%%%%%%%%%%%%%%%%%%%%%%%%%%%%%%%%%%%%%%%%%%%%%%%%%%%%%%%%%%%%%%
%%%%%%%%%%%%%%%%%%%%%%%%%%%%%%%%%%%%%%%%%%%%%%%%%%%%%%%%%%%%%%%%%%%%%%
%%%%%%%%%%%%%%%%%%%%%%%%%%%%%%%%%%%%%%%%%%%%%%%%%%%%%%%%%%%%%%%%%%%%%%
%%%%%%%%%%%%%%%%%%%%%%%%%%%%%%%%%%%%%%%%%%%%%%%%%%%%%%%%%%%%%%%%%%%%%%

\end{document}